\documentclass[nonacm,sigconf]{acmart}

\def\BibTeX{{\rm B\kern-.05em{\sc i\kern-.025em b}\kern-.08emT\kern-.1667em\lower.7ex\hbox{E}\kern-.125emX}}

\usepackage{hyperref}
\usepackage{tikz}
\usepackage{soul}
\usepackage[nohyperlinks, nolist]{acronym}
\usepackage{tabularx}
\usepackage{enumitem}
\usepackage{amsmath}
\usepackage{cleveref}
\usepackage{xspace}
\usepackage{calc}
\usepackage{pzccal} 

\newcolumntype{Y}{>{\centering\arraybackslash}X}

\crefname{appsec}{Appendix}{Appendices}

\renewcommand\footnotetextcopyrightpermission[1]{} 

\newcommand{\sysname}{Grand\-Det\-Auto\xspace}
\newcommand{\advshare}{$\alpha$\xspace}

\newcommand{\node}[1]{$n_{#1}$\xspace}
\newcommand{\adv}{$\mathpzc{A}$\xspace}
\newcommand{\waitcert}{$w$\xspace}

\begin{acronym}
	\acro{bft}[BFT]{Byzantine Fault Tolerance}
	\acro{pbft}[PBFT]{Practical Byzantine Fault Tolerance}
	\acro{diat}[DIAT]{Data Integrity ATtestation}
	\acro{mac}[MAC]{Message Authentication Code}
	\acro{tcb}[TCB]{Trusted Computing Base}
	\acro{tpm}[TPM]{Trusted Platform Module}
	\acro{tee}[TEE]{Trusted Execution Environment}
	\acro{rop}[ROP]{Return-Oriented Programming}
	\acro{poet}[PoET]{Proof-of-Elapsed-Time}
\end{acronym}

\begin{document}

\date{}

\title{GrandDetAuto: Detecting Malicious Nodes in Large-Scale Autonomous Networks}

\author{Tigist Abera}
\affiliation{%
	\institution{Technical University Darmstadt}
	\city{Darmstadt}
	\country{Germany}
}
\email{tigist.abera@trust.tu-darmstadt.de}

\author{Ferdinand Brasser}
\affiliation{%
	\institution{Technical University Darmstadt}
	\city{Darmstadt}
	\country{Germany}
}
\email{ferdinand.brasser@trust.tu-darmstadt.de}

\author{Lachlan Gunn}
\affiliation{%
	\institution{Aalto University}
	\city{Greater Helsinki}
	\country{Finland}
}
\email{lachlan.gunn@aalto.fi}

\author{Patrick Jauernig}
\affiliation{%
	\institution{Technical University Darmstadt}
	\city{Darmstadt}
	\country{Germany}
}
\email{patrick.jauernig@trust.tu-darmstadt.de}

\author{David Koisser}
\affiliation{%
	\institution{Technical University Darmstadt}
	\city{Darmstadt}
	\country{Germany}
}
\email{david.koisser@trust.tu-darmstadt.de}

\author{Ahmad-Reza Sadeghi}
\affiliation{%
	\institution{Technical University Darmstadt}
	\city{Darmstadt}
	\country{Germany}
}
\email{ahmad.sadeghi@trust.tu-darmstadt.de}

\begin{abstract}
Autonomous collaborative networks of devices are rapidly emerging in numerous domains, such as self-driving cars, smart factories, critical infrastructure, and Internet of Things in general.
Although autonomy and self-organization are highly desired properties, they increase vulnerability to attacks. 
Hence, autonomous networks need dependable mechanisms to detect malicious devices in order to prevent compromise of the entire network.
However, current mechanisms to detect malicious devices either require a trusted central entity or scale poorly.

In this paper, we present \sysname, the first scheme to identify malicious devices efficiently within large autonomous networks of collaborating entities. 
\sysname functions without relying on a central trusted entity, works reliably for very large networks of devices, and is adaptable to a wide range of application scenarios thanks to interchangeable components.
Our scheme uses random elections to embed integrity validation schemes in distributed consensus, providing a solution supporting tens of thousands of devices.
We implemented and evaluated a concrete instance of \sysname on a network of embedded devices and conducted large-scale network simulations with up to $100\,000$ nodes. 
Our results show the effectiveness and efficiency of our scheme, revealing logarithmic growth in run-time and message complexity with increasing network size.
Moreover, we provide an extensive evaluation of key parameters showing that \sysname is applicable to many scenarios with diverse requirements.
\end{abstract}

\maketitle


\section{Introduction}
\label{sec:introduction}
The growing trend towards the Internet of Things (IoT) and Autonomous Systems allows connected devices to collaborate, enabling more efficient as well as new applications.
This opens up new opportunities in many domains, from self-driving cars and smart factories to critical infrastructure.
Various industries are motivated by higher efficiency and increased flexibility, which can be achieved by connecting devices within individual factories as well as by interconnecting facilities collaborating within a supply-chain~\cite{SOM14}.
Other industry branches, like the automotive and associated industries, strive to increase safety through connection and collaboration, e.g., cars sharing information about potential hazards~\cite{JH11}.
The extensive efforts to standardize vehicle communications by major industry leaders shows the relevance of this trend, such as the cellular network-based \mbox{C-V2X}~\cite{c_v2x} and the WiFi extension standard 802.11p~\cite{v2v_wifi}, which, for instance, Volkswagen announced to support in \emph{all} 2020 Golf 8~\cite{golf_c2x}.
However, despite these advantages autonomous systems bear various security risks. 
Hence, it is important to develop solutions for these challenging scenarios.
In particular, in safety-related scenarios, malicious devices can cause tremendous damage and threaten human life.

To secure such systems many proposals rely on a central authority~\cite{seda,sana,darpa,lisa,ids_central,ids_central2}. 
However, a centralized solution constitutes a single point of failure, implying unrealistic requirements on the central authority:
(1)~The availability of the authority must be guaranteed at all times, i.e., the entire system must have continuous and reliable connectivity to it.
This is hard to guarantee in many practical systems, e.g., with freely moving nodes.
(2)~A central authority is an attractive attack target, exposing it to a wide range of attacks.
Any successful attack will corrupt its integrity and/or availability, i.e., make the central authority fail.
There are many real world examples how centralization of authority can be detrimental, like the compromised DigiNotar PKI (public key infrastructure) issuing fraudulent certificates for Google, Microsoft and CIA websites~\cite{ca_diginotar}, the attack on Ukraine's power grid by compromising centrally operated Industrial Control Systems~\cite{ca_blackenergy}, or the DDoS attack on DynDNS bringing down major websites (incl. PayPal, CNN and Amazon) in parts of Europe and the US~\cite{ca_mirai}.
These examples show that even the most sophisticated defense mechanisms aiming to protect central services can be circumvented.
Further, when multiple (mutually distrusting) stakeholders are involved, it is difficult to jointly agree on a party that acts as the trusted authority. 
For instance, different car manufacturers or cellular network equipment providers, which in many cases do not inherently trust each other, will not easily agree on an overarching authority with the power to control all devices.
\\
\noindent\\
\textbf{Problem.}
Strongly reducing or fully eliminating the role of the central party in connected systems seems very appealing; yet, it requires the connected devices to collaborate and share information in a broadly autonomous fashion.
Consequently, interdependencies within the network will increase the threat that malicious devices could pose on the entire system. 
Increasingly interconnected devices, including modern vehicles~\cite{KCR+10,CMK+11,MV14}, industrial facilities~\cite{stuxnet,duqu,BL04}, critical infrastructure~\cite{blackenergy,Kabay10,PC10}, and even medical devices~\cite{insulin} have been the targets of attacks. 
In particular, a single malicious device could cause other devices to deviate from the correct behavior; for instance, influencing the routing of other cars by transmitting false traffic information~\cite{SPY+15}. 
Hence, large autonomous networks must also be able to identify faulty or malicious devices in order to react to attacks.
It is paramount to prevent a (partial) compromise of the network from impairing the correct function of the overall system.
\\
\noindent\\
\textbf{Existing defense strategies.} 
Attack \emph{detection} methods can uncover ongoing attacks, enabling more sophisticated reaction policies, like the recovery of a compromised device~\cite{polygraph} to prevent an adversary taking over the network. %
Outlier detection is used in Wireless Sensor Networks 
to identify outliers on aggregated sensor data, which may be caused by malicious attacks~\cite{wsn_survey}; yet, many directly rely on a central entity.
Thus, they are inapplicable to autonomous systems without central authority.
Approaches that do not rely on such a central entity~\cite{wsn_neigbors,wsn_vote} do not scale for large networks commonly encountered in autonomous systems.
There are collaborative intrusion detection approaches that distribute data acquisition across the network~\cite{ids_adhoc_surv}.
Yet, they either assume a central authority for decision-making~\cite{ids_central,ids_central2}, or assume only few, sparsely distributed malicious devices in the network~\cite{ids_majority_manet,ids_cluster1,ids_cluster2}.
For real-world scenarios, this is hard to guarantee as adversarial nodes can collaborate to gain the majority in a group of nodes.
Other approaches, such as swarm attestation~\cite{seda,sana,darpa,lisa},
provide an integrity proof for the whole network to a central verifying entity.
Thus, they are inapplicable to autonomous systems without central authority.
We elaborate more on these approaches in \Cref{sec:related}.

\noindent\\
\textbf{Goals and Challenges.}
Designing an efficient scheme for detecting and identifying malicious nodes/devices in a connected autonomous system faces us with a number of challenges. 
The overarching challenge is scalability: A naive solution in which each node individually performs monitoring and validation of potentially all other nodes
is inefficient, especially with resource-constrained embedded devices.
Thus, the naive solution does not scale.
Instead, an appropriate scheme needs to combine local monitoring with efficient and scalable decision-making.
For this, we derive three key challenges.

\noindent
\textbf{Challenge 1: Flexible and adaptive detection of malicious devices.}
To be able to identify compromised devices, a practical mechanism is needed to validate whether a device is in a good state, i.e., behaving as expected. 
There are various approaches to achieve this, each coming with a set of advantages and disadvantages. 
We discuss this in \Cref{sec:framework:verification}.

\noindent
\textbf{Challenge 2: Establishment of a network-wide shared state.}
In addition to having a scheme for device state integrity validation, 
a common state among the nodes is needed.
However, agreeing on a common state efficiently among all individual nodes is particularly difficult in large-scale networks.

\noindent
\textbf{Challenge 3: Resiliency.}
In an autonomous system, the monitoring as well as the decision-making is generally distributed among the nodes.
To guarantee resiliency, the final decision must not only rely on monitoring results raised by an individual node, but on a distributed agreement.
However, a distributed decision-making scheme must be carefully 
designed to avoid introducing new attack vectors.

\vspace{0.1em}
\noindent\\
\textbf{Contributions.}
In this paper, we present \sysname, a novel distributed adversary detection scheme for large-scale networks. 
The design of \sysname is generic and modular.
It combines schemes for integrity validation of devices' states with schemes for distributed election and consensus in a novel way, while each of these modules can be instantiated with different primitives that fit the requirements posed by the corresponding application.
More precisely, any device may blame another device for being malicious by providing a proof that the state integrity of that device is violated. 
This proof will then be verified by a randomly and autonomously selected jury (a subset of devices), which in turn finds a consensus on whether the proof is valid.
Because the jury-size is fixed but configurable, the consensus overhead remains constant independent of network size (aside routing).

Our main contributions include:

\begin{itemize}
	\item \sysname is the first efficient and dependable scheme to allow a system of collaborating entities without a central authority to detect its compromised parts by distributing integrity validation schemes via random elections leading to Byzantine fault-tolerant decisions (\Cref{sec:design}).
	\item \sysname is highly flexible since its components can be instantiated by various schemes for integrity validation, random elections, and consensus protocols (\Cref{sec:framework}).	
	\item We introduce a novel distributed election scheme, inspired by Proof-of-Elapsed-Time~\cite{poet}, to randomly elect a \emph{group} of representatives in the network.
	\item We implemented a \sysname prototype in the context of smart traffic based on the aforementioned election scheme, \ac{pbft}~\cite{pbft} and remote attestation (\Cref{sec:impl}) using an ARM platform with TrustZone.
	\item Being a distributed system, \sysname's efficiency and security relies on a suitable choice of key parameters, which we thoroughly analyze and evaluate (\Cref{sec:eval}).
	Further, we developed a large-scale network simulation with tens of thousands of devices and demonstrate \sysname's scalability through extensive evaluation (\Cref{sec:eval_perf}). 
\end{itemize}


\section{System Model} \label{sec:model}
\acresetall
We consider large distributed autonomous systems; specifically, a network of connected devices \node{1}, ..., \node{i} that collaborate with each other to perform complex tasks. 
We use the terms device or node interchangeably in the following.
Nodes may join and leave the network; yet, the list of devices participating in the network is known\footnote{Managing membership is an orthogonal problem with existing solutions; we outline one in Section~{\ref{sec:impl:election}}.}
In order to collaborate by coordinating their actions, the individual entities of the overall system need to exchange information, such as status updates and sensor readings, which is often critical for the correct behavior of the overall system.
In a smart traffic scenario, for example, false position information may lead to vehicles crashing into each other.

All devices are mutually distrusting and there is no trusted central entity or external coordinating operator on which the network must rely.
\sysname has a generic design and does not assume any specific security framework or security hardware. However, depending on the instantiation in practice, it can utilize security architectures, such as \acp{tee} for random election and integrity validation, as we present in Section \Cref{sec:impl}.

\subsection{Adversary Model and Assumptions} \label{sec:advmodel}
The adversary's goal is to influence the collaboration between honest nodes by manipulating the data sent to other devices.
We make the following assumptions about the adversary's capabilities.
The adversary \adv is able to compromise and coordinate a subset of devices in the system.
We denote \advshare as the threshold of malicious devices our scheme can endure.
\advshare depends on the system parameters, which we discuss and extensively evaluate in \Cref{sec:probs}.
Compromising new devices takes non-negligible time for the adversary\footnote{Assuming basic security like memory layout randomization, exploiting devices requires many attempts~\cite{PaX-ASLR,aslp,WMH+12,PPK12,ilr,readactor,lr2}.}.
We further assume that the adversary cannot break cryptographic primitives.
Devices that participate in denial-of-service (DoS) attacks are considered malicious in our system\footnote{As a result those devices will be handled by the recovery mechanism, e.g., by expelling them.}.

We assume \adv can eavesdrop and manipulate messages between devices.
However, \adv is limited to disturbing the communication of nodes within physical proximity, e.g., via jamming.
Hence, \adv can control only a subset of all network links, preventing it to block overall communication in the network.
This can be realized through various network technologies, e.g., meshed networks with robust routing~\cite{GMF14,MP14}, or upcoming technologies like 5G~\cite{5g} and satellite-based networks~\cite{starlink, oneweb} where malicious network-clients have very limited means to disturb the overall network communication.
Finally, \adv isolating individual devices can be inherently tolerated by \sysname as faults in the consensus phase.

In addition, for a concrete instantiation we inherit the security guarantees and assumptions of the components used by \sysname.
For instance, if we use remote attestation to validate the software state of a device, the respective assumptions of the remote attestation framework will apply to \sysname. 
This means that we may assume the existence of some trust anchor on the involved devices and consider physical attacks out of scope. 
Similarly, \sysname inherits the protection capabilities of the used components, e.g., different attestation schemes can detect different types of software attacks.

\subsection{Requirements}
\label{sec:requ}
A scalable and flexible malicious device detection scheme for collaborative autonomous networks shall fulfill the following properties:

\begin{enumerate}[nolistsep, label=\textbf{R.\arabic*:}, ref={R.\arabic*}, labelsep=\widthof{~}, itemindent=0pt] 
	\item \label{req:det_ident}\emph{Detection and Identification:} On the one hand, an adversary trying to maliciously interfere with the network shall be detected.
	On the other hand, if the adversary tries to manipulate the overall scheme at any point, it shall be detected as well.
	\item \label{req:efficieny}\emph{Efficiency:} Validating the integrity of a device must be significantly more efficient than letting all nodes validate that device individually.
	\item \label{req:scalability}\emph{Scalability:} The computational effort and communication complexity grows sub-linear with respect to the number of devices (scaling to large networks). 
	\item \label{req:interchangeable}\emph{Interchangeable Components:} Individual components have clearly separated roles and objectives, making them easily replaceable.
\end{enumerate}


\begin{figure}[ht]
	\centering
	\includegraphics[width=0.8\columnwidth,trim=0cm 0.4cm 2.6cm 0.2cm, clip]{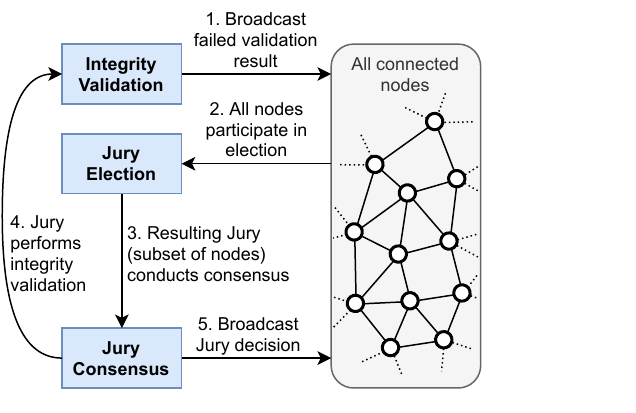}
	\caption{Interactions between the components.}
	\label{fig:approach_components}
\end{figure}

\section{\sysname Design} \label{sec:design}

\sysname provides a scalable solution to detect malicious devices in truly autonomous networks, i.e., without external supervision from a central entity.
It consists of three main components, as seen in \Cref{fig:approach_components}, and works as follows.
A node uses \emph{Integrity Validation} on another node and if this validation fails, the node will announce the other node as suspicious by broadcasting the validation result (1. in the figure).
This starts the second phase of the protocol.
In this phase, the \emph{Jury Election} will select a group of nodes acting on behalf of the whole network, i.e., the jury (2.).
The resulting jury will then use the \emph{Jury Consensus} to reach an agreement (3.) by confirming the initial integrity validation (4.).
Finally, after the jury reached a consensus, the decision will be broadcast to the rest of the network (5.).
This jury decision can enforce an action, e.g., excluding a malicious node from the network.
We discuss this aspect in \Cref{sec:discussion}, which is not in scope of this work.

\begin{figure}[ht]
	\centering
	\includegraphics[width=\columnwidth,trim=0 0.2cm 0 0, clip]{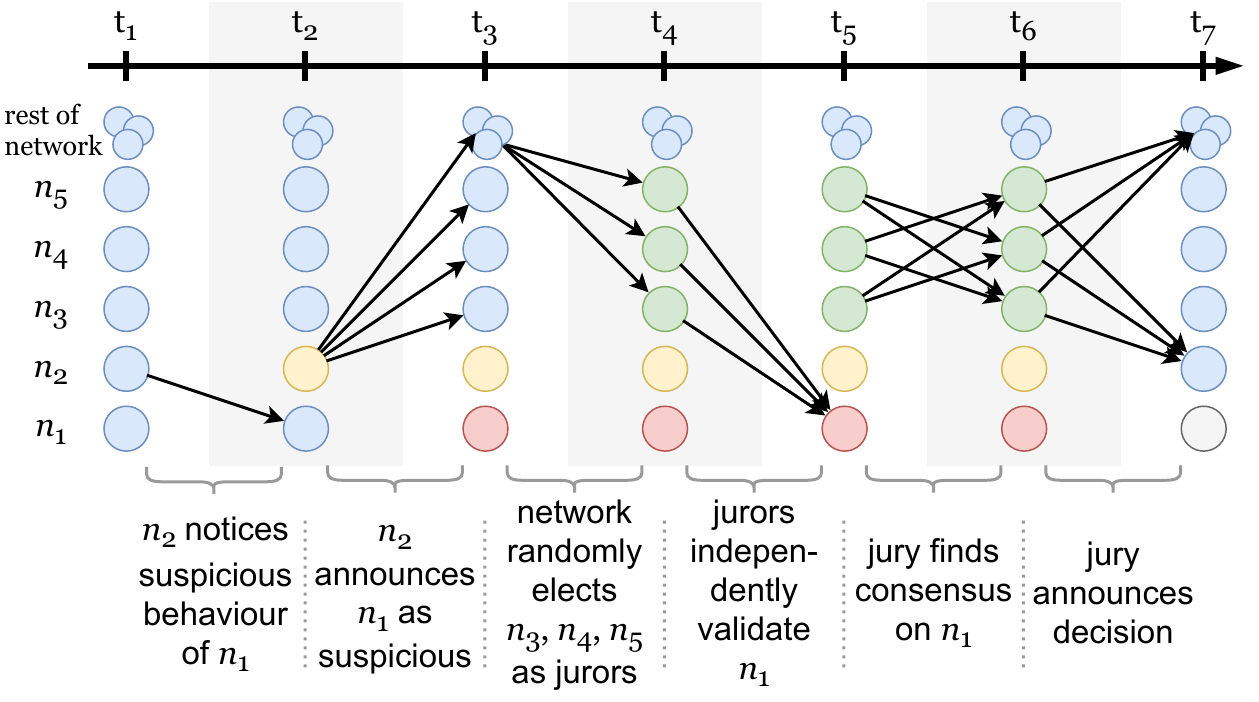}
	\caption{An exemplary setup with one adversary (red) and three jurors (green), \emph{rest of network} refers to \node{6}, ..., \node{i}.}
	\label{fig:approach_overview}
\end{figure}

\Cref{fig:approach_overview} illustrates an exemplary run of \sysname. 
After \node{2} notices \node{1} suspicious behavior, we call the announcement of this suspicion to the rest of the network \emph{blaming}.
Next, the network randomly elects the jury in a distributed manner, in this case \node{3}, \node{4}, \node{5}.
Each juror will individually validate the claim made by \node{2}, find a consensus about the blamed \node{1} as well as the decision how the network shall react, and broadcast the result among the network.
This example solely illustrates one round of \sysname, i.e., one processed suspicion.

\sysname is triggered on-demand by individual nodes, and thus does not entail any overhead in a network of benignly acting nodes.
Instead, all nodes individually look for suspicious behavior among local nodes that are potentially malicious.
Depending on the scenario in which \sysname is used, suspicious behavior can be detected in various ways.
For instance, inconsistencies in sensor readings exchanged between devices can be used to detect adversarial devices, as demonstrated in industrial control systems~\cite{ECK+17, CBI}.
Similar approaches could be used in settings like smart traffic, where different cars can match their sensor readings against reported data from other cars to detect inconsistencies.
Suspiciously acting nodes are then examined thoroughly. 
We achieve this by giving each node the ability to \emph{blame} another node, i.e., broadcasting to the whole network that a node acts suspiciously and may be adversarial. 
This approach drastically limits the adversary's impact on the overall system.
As soon as the adversary starts to impact other devices, e.g., by sending false information, it will be quickly detected and sanctioned.
In \Cref{sec:discussion} we also discuss a modification for \sysname to detect passive adversaries as well.

Once a node is blamed, the network has to reach a decision whether the node is malicious.
For a sustainable autonomous network, it is important to have a consistent view across the network including the order of processed blames, as concurrent ones can conflict. 
Thus, a form of consensus is needed. 
However, consensus protocols do not scale to large networks, due to their exponential message complexity~\cite{pbft}. 

To overcome this fundamental limitation, \sysname randomly elects a jury as the representatives to make a decision for the whole network.
Ensuring a fair election in a distributed system is important for the security of the overall system, as the mutually distrusting nodes need to reliably agree on a common jury.
Due to the distributed nature of \sysname, the challenge is to reliably converge on an election result.
Otherwise, the individual views on which nodes are part of the jury may diverge and cause additional faults for the consensus or at worst, entirely prevent finding a quorum.
Therefore, the right choice of key parameters guiding the election is critical.
We examine the effect of such parameters in \Cref{sec:eval_waittime} and demonstrate a suitable trade-off between election reliability and the run-time of \sysname.

While executing consensus only on a subset of nodes improves scalability, it comes at the expense of the consensus' safety.
As the election of the jury is random, there is a chance that a sufficient number of malicious nodes are among the elected jury so that they can enforce an adversarial decision within the jury.  
However, we can adjust the consensus so that it stalls rather than fails, as stalling can be rectified by a re-election.
In \Cref{sec:probs} we analyze how these probabilities behave regarding \sysname's configurable parameters. 
We show that these parameters can be chosen such that the probability of electing an adversarial jury is negligible.

\noindent
\textbf{Resiliency.}
A challenging problem to address is how the system can defend itself against abuse.
More specifically, an adversarial node may try to blame an honest node to disrupt the system, e.g., blaming the blamer.
Furthermore, the adversary may try this multiple times to increase the chances of electing enough accomplices to successfully seize the jury, or simply try to use the blaming mechanism to overload the system with requests.
In case a blame was unjustified, the jury will decide to blame the potentially dishonest blamer, immediately starting another round to determine if the blamer is indeed malicious.
Further, a node clearly violating the expected behavior of an underlying component can result in the node being blamed as well.
For example, when the validation process is deterministic, correct jurors can safely blame a juror that reaches a different conclusion from the same data.
These automatic blaming approaches will prevent the adversarial nodes to turn the chances in their favor over time, as attempts to manipulate the protocol will in turn risk getting blamed themselves.


\section{\sysname Design Decision} \label{sec:framework}
\sysname is designed to be modular; hence, individual components for each phase (integrity validation, random jury election and consensus) can be instantiated differently, based on the requirements of the underlying application; thus, \sysname fulfills~\ref{req:interchangeable}. 
This section will enumerate the options we identified for each component and state the choices made for our instantiation of \sysname we present in \Cref{sec:impl}, which also further elaborates on the chosen schemes.

\noindent
\paragraph{Integrity Validation Scheme.}\label{sec:framework:verification}
\sysname requires a mechanism for detecting the initial suspicious behavior of a potentially malicious device as well as a mechanism for the jurors to validate blames. 
More concretely, it should be possible to verify the integrity of a node, whether its behavior or state deviates from what is expected.
The integrity validation should not rely on a central trusted entity and be able to run on devices with limited computational resources.
Further, the result of the validation (e.g., through a node) should be verifiable by other devices.

There is a rich body of literature on proposals to determine whether a device is behaving as expected.  
We identified the following options: Unsupervised outlier detection for sensor data \cite{wsn_survey,wsn_vote,wsn_neigbors}, which is the prevalent method used in Wireless Sensor Networks; Intrusion Detection Systems (IDS) monitor for anomalies in network traffic in order to discover intrusions; or Remote attestation, which is a security primitive that enables a verifying party to receive direct proof that the software of a remote device is in a trustworthy state based on the verifier's trust policy (e.g., the code is not altered).
We leverage remote attestation to instantiate the integrity validation component of \sysname, as this approach is agnostic towards the targeted use case, opposed to the careful consideration required to define outliers or anomalies, which are highly context-specific.
As the node's program (execution) intrinsically defines its behavior, attestation can detect maliciously acting nodes.
In Section \Cref{sec:options:validation} we will elaborate on the aforementioned approaches as well.

Note, it is possible to use two distinct validation schemes for different phases of \sysname. 
As shown in \Cref{fig:approach_overview}, the initial validation raising the suspicion can be done with a lightweight but overestimating scheme like outlier detection, e.g., by observing inconsistencies in communication with another party.
Then the elected jury can perform a thorough and complex scheme like remote attestation to validate this initial suspicion.

\noindent
\paragraph{Random Jury Election.} \label{sec:election}
After a node has been blamed, the network randomly elects a jury. 
For \sysname, the election scheme should work in a distributed and verifiable manner as well as ensure fairness, i.e., every node has the same chance of being elected.
Approaches for distributed random elections can be found in the blockchain space.
Their goal is to elect the proposer for the next block by a fair ``lottery''.
Their security is usually based on monetary incentives to prevent Sybil attacks~\cite{sybil}, i.e., a node assuming multiple identities to unfairly increase its influence.
Unfortunately, this means they are not directly applicable for our purpose.

However, we identified the following non-incentivized schemes:
Algorand~\cite{algorand} elects a delegation group to propose the next block by leveraging a Verifiable Randomness Function; Byzcoin~\cite{byzcoin} also uses a delegation for block proposal based on their success mining blocks via a Proof-of-Work scheme; or Intel's \ac{poet}~\cite{poet}, which forces nodes to wait for a random amount of time and the ``fastest'' node may propose a block.
Most relevant for our instantiation of \sysname is \ac{poet}, as it can significantly reduce message overhead for the network (see \Cref{sec:impl:election}).
However, as it is designed to elect a single node, we extend the scheme to be able to elect a group of nodes, i.e. the jury, as described in \Cref{sec:impl:election}.
Section \Cref{sec:options:election} will discuss the other mentioned schemes as well.

\noindent
\paragraph{Consensus} \label{sec:framework:consensus}
After all jurors performed their individual integrity validation of the blamed node, they need a consensus scheme to agree on the result and the reaction to it. 
Keeping a consistent order of the jury decisions is crucial, as multiple simultaneous blame requests may occur that depend on each other. 
For example, one round may elect a juror that is expelled from the network in another round.
In \sysname, a consensus scheme should ensure that blame requests are consistently processed, including their order.

While this can be achieved via the inherent properties of the election (see \Cref{sec:options:consensus}), the use of a consensus protocol, i.e., \ac{bft}, eliminates the need to do an election on every blame, significantly reducing the overhead of the elections over multiple rounds.
This way, we can keep an elected jury for a selectable time window.
Especially if multiple nodes are blamed in quick succession, \ac{bft} can have a significantly higher throughput.
We use \ac{pbft}~\cite{pbft} for our instance of \sysname.
There are variations of \ac{pbft} that may also be used, e.g., to improve performance if malicious behavior is expected to be rare.
We elaborate on these alternatives in \Cref{sec:options:consensus}.
However, the de-facto baseline in the \ac{bft} literature is \ac{pbft}~\cite{pbft} and does not introduce additional assumptions.


\section{Implementation}
\label{sec:impl}
\label{sec:impl2}
\begin{figure}[ht]
	\centering
	\includegraphics[width=\columnwidth,trim=0cm 0.3cm 0cm 0.1cm, clip]{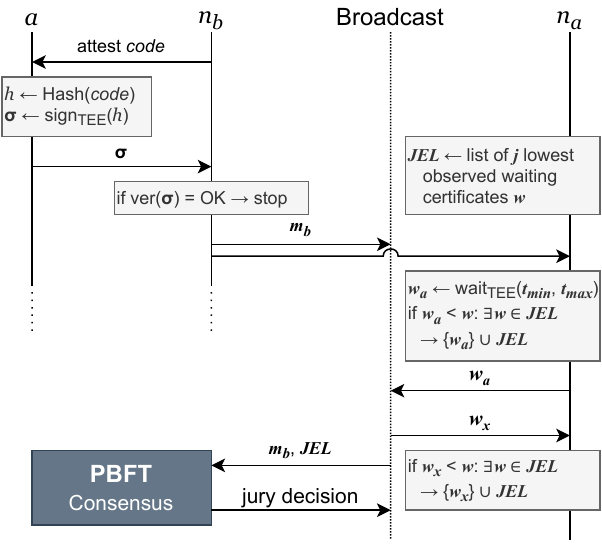}
	\caption{Overview of the protocol instance.}
	\label{fig:impl}
\end{figure}

Subsequently, we present our full implementation of an instance of \sysname for a smart-traffic use-case using off-the-shelf devices.
Our prototype uses an ultrasonic sensor to perform distance measurements that are shared with other road users. 
This is a common task in smart traffic scenarios~\cite{smarttraffic}, in which sharing environment-sensing data is crucial for vehicles to avoid collisions.
\sysname can be used in this scenario to identify adversarial vehicles that send altered environment-sensing data, and endanger other vehicles.
A neighboring node raises suspicion whenever the measured distance changes abruptly (assuming the measurement can be modeled as a continuous function), leading to an initial attestation by the neighbor.
This starts a round of \sysname.
\Cref{fig:impl} shows an overview of the communication flow.

As the underlying platform we used ARM TrustZone~\cite{trustzone} running Open Portable Trusted Execution Environment (OP-TEE) OS \cite{optee}, in line with previous works for automotive use cases using TrustZone~\cite{hu2020cvshield, lee2019t, liu2017protc}.
TrustZone divides the system into a secure world and a normal world by loading a trusted OS separately from the normal OS, e.g., Linux.
Sensitive code is executed in the secure world, in so-called \emph{Trusted Applications (TAs)}.
TAs may only communicate with the normal world with explicitly allocated shared memory regions.
TAs are managed by the secure-world OS, in our case OP-TEE OS.
The main protocol, handling the communication and the Byzantine agreement, as well as the distance measurement software run as normal-world applications, while the attestation and the random election are implemented as TAs.
These small TAs communicate with the main protocol executing in the normal world.
In the following, we describe each of the implemented components in detail.

\subsection{Device Attestation} \label{sec:impl:att}
To validate the integrity of a device, remotely attesting its software can provide strong security guarantees, as it enables nodes to directly prove that their software is not altered.
There are different attestation schemes, which can detect different classes of software attacks~\cite{FNR+14,SLP08,cflat,litehax,diat}.
Our implementation uses binary remote attestation as the integrity validation scheme.
Binary attestation is still a relatively simple mechanism, and hence is not secure against software run-time attacks using techniques such as code-reuse attacks~\cite{FC08} that do not require any modification to the code leaving the hash value unchanged.
However, any other (e.g., more sophisticated) attestation scheme can be used as well, such as (hardware-assisted) run-time attestation schemes that can also be used to detect run-time modifications. 
We instantiate the integrity verification component with a more enhanced attestation scheme for evaluation in \Cref{sec:impl_hw_eval}.

Our binary attestation scheme enables \sysname to detect any unintended modification of the code, as such modifications are recorded and included in the attestation report $\sigma$.
In \sysname, if a verifier node's attestation of another node is negative, it can use the attestation report as the evidence for the jury to confirm its blame.
The attestation process is implemented using a TA in TrustZone's secure world based on OP-TEE.
The attestation TA computes the hash of the normal-world app, i.e., its \emph{code} in memory, and signs it with a signing key, which is kept confidential inside the TA.
The signing key needs to be issued by the device vendor and must be part of a public key infrastructure such that other nodes can verify signed attestation reports.
This signed attestation report $\sigma$ is then sent to the verifier, i.e., another device.
If the verifier finds that the state reported in $\sigma$ is not trustworthy, it blames the prover device by broadcasting the \emph{blame message} $m_{b}$ containing $\sigma$.

\subsection{Random Jury Election} \label{sec:impl:election}
For the jury election, we use an approach inspired by Intel's \ac{poet} for the random jury election.
\ac{poet} leverages \acp{tee} and a registration process based on linkable attestation, i.e., attestation directly bound to a specific processor, to prevent Sybil attacks.
Each participant gets a publicly verifiable random number and needs to wait for this random amount of time instead.
For this feature the \ac{tee} is used as well, which attests that the respective node has indeed waited for its assigned amount of time, i.e., generating a \emph{waiting certificate}.
Afterwards, the participant will broadcast this result and each node will deem the earliest observed waiting certificate as the election's winner.
This approach inherits a degree of fault-tolerance, as a crashed election winner can easily be replaced with the next best node.
Further, the waiting approach reduces message complexity. 
Honest nodes with a comparatively high number will also wait longer and may observe lower-valued waiting certificates in the meanwhile. 
In this case, the node will decide not to announce its own wait certificate, saving overhead as only a minor part of the network needs to announce their respective wait certificates. 

However, this approach is designed to only elect a single block proposer.
Thus, we designed and implemented our own approach to elect a group of nodes, i.e., the jury.
To implement a waiting approach analogously to \ac{poet}, we use OP-TEE functions providing a secure timer (\texttt{TEE\_GetSystemTime}) as well as a secure wait function (\texttt{TEE\_Wait}).
Both are used to implement the wait TA.
As the wait time is derived from publicly known data, the wait time itself is publicly verifiable.
Thus, the node simply passes the wait time to the wait TA, which waits for the given amount of time.
By getting the trusted system time inside the TEE before and after the wait, we simply have to sign both timestamps to get a valid waiting certificate \waitcert.
This procedure is done on every node in the network.

The interplay between the different nodes during the election of $j$ jurors works as follows:
\begin{enumerate}
	\item As soon as a node receives a blame message $m_{b}$, it will generate a random waiting time chosen from a exponential distribution and wait for the generated amount of time. 
	The waiting time is in the range between a defined minimum $t_{min}$ and  maximum $t_{max}$. 
	\item Each Node will receive waiting certificates from other nodes to compile a Jury Election Leaderboard $JEL$, a sorted list containing waiting certificates with the lowest observed waiting times. 
	When a node \node{a} receives a waiting certificate from another node $w_{x}$, the node will first check if the certificate has merit, i.e., if the waiting time is lower than the largest entry in $JEL$ or if $|JEL|<j$. 
	Nodes refrain from validating and forwarding any \waitcert if they do not have merit at that time. 
	Otherwise, the node will check the validity of $w_{x}$ and add it at the corresponding position in its $JEL$. 
	If $|JEL|>j$, the last entry is removed so $|JEL|=j$.
	
	\item After \node{a} is finished waiting, it will check if $|JEL|<j$ or if its own waiting time is smaller than any of the entries in $JEL$. 
	If its own waiting certificate $w_{a}$ has merit, it will announce it to the network and add it to its $JEL$. 
	If not, it will discard $w_{a}$.
	
	\item After \node{a} additionally waited for a pre-defined time threshold $t_{ele}$, it will assume its $JEL$ to be mostly complete. 
	If the node's own $w_{a}$ is still in $JEL$, it will assume to be part of the jury. 
	If so, it will start the agreement process with the other jurors found in $JEL$.
\end{enumerate}

This way, all nodes will eventually converge towards an identical Jury Election Leaderboard $JEL$ consisting of the $j$ waiting certificates with the smallest waiting time.
Thus, it is essential to choose suitable values for $t_{min}$, $t_{max}$, and $t_{ele}$ to ensure a reliable election.
We extensively evaluate these parameter in \Cref{sec:eval_waittime}.

\subsection{Byzantine Agreement} \label{sec:bftimpl}
To find a consensus among the jurors about a blamed node, we chose to implement \ac{pbft}~\cite{pbft}. 
Using a \ac{bft} scheme eliminates the need to do an election on every blame, significantly reducing the overhead of the elections over multiple rounds. 
Especially if multiple nodes are blamed in quick succession, \ac{bft} can have a higher throughput.

In general, \ac{bft} is used to find a consensus among a group where some might be faulty or adversarial, i.e., act \emph{Byzantine}.
In \ac{bft} at least $3f+1$ total nodes are required to endure $f$ Byzantine nodes~\cite{bft}.
Traditional \ac{bft} schemes assume all nodes participate in the agreement process; thus, 
if the process fails due to too many Byzantine nodes, it is impossible to succeed.
However, in our case, every election will have a diverse agreement group and may succeed where the previous jury failed. 
A failed \ac{bft} agreement does not prevent progress in \sysname, as the failure can trigger a new election resulting in a new jury that is likely to proceed.
Furthermore, we consequently need to consider an additional negative case that we label the \emph{total fail} case.
When enough adversarial jurors are elected to reach a quorum, they can enforce a malicious decision, i.e., violating \ac{bft}'s \emph{safety} guarantee.
Further, the consensus can also fail due to not reaching a quorum at all, i.e., violating the \emph{liveness} guarantee.
In \Cref{sec:probs} we examine the probabilities of both re-elections and total fails.

In \ac{pbft}, the primary decides on which request is being processed next by the consensus group.
In the \emph{prepare} phase, the participants exchange this request among each other to ensure the primary sent all participants the same request.
Afterwards, the participants execute the request and exchange their results in the \emph{commit} phase.

\noindent
Our scheme works as follows, using \ac{pbft} as a subprotocol:

\begin{enumerate}
	\item After the election, the Jury Election Leaderboard $JEL$ has a sorted list of the $j$ lowest wait times for each juror. 
	The juror with the lowest wait time will be the primary.
	\item On conflicting blame requests and elections, the jury containing the overall shortest waiting time is selected for the next round. 
	Thus, the initial round for a new jury can skip the prepare phase entirely.
	\item Otherwise, \ac{pbft} will be executed among the jury about the validity of $\sigma$ included in the agreed on $m_b$.
	\item The jury decision (see \Cref{sec:discussion}) is then broadcast by all jurors, containing at least two thirds of all jurors' signatures.
	The rest of the network can consider each valid and consistent decision message on the same blame to be equivalent. 
	This avoids separately spreading up to $j$ inconsequentially different decision messages.
\end{enumerate}

\subsection{Communication Aspects}
In our implementation, two aspects with regards to communication are particularly relevant: 
(i) Broadcasting was implemented by using a flooding-based protocol. 
Every node forwards broadcast messages to all neighbors, except the one from which the message was originally received. 
This way, a message will take the optimal paths, and thus flooding is optimal regarding run-time.
We discuss alternatives in \Cref{sec:discussion}.
(ii) To reduce communication overhead in terms of message sizes (in bytes), we use a collective signature scheme. 
In the consensus phase, all jurors have to individually consent by providing their own signatures. 
As we evaluate different jury sizes, we decided to implement the Schnorr signature scheme~\cite{collsig}. 
This way, increasingly adding signatures to a message does not result in increasing \ac{bft} messages sizes.


\section{Security Evaluation} \label{sec:eval}
\acresetall
In this section, we evaluate \sysname's security and present an analysis the probabilities of the jury consensus to fail.
We will show how different parameters affect \sysname and provide the foundation for selecting a suitable configuration. 

\subsection{Security Consideration} \label{sec:security}
As mentioned in the Sections \ref{sec:advmodel} and \ref{sec:requ}, the adversary's goal is to either evade being identified (detected) by \sysname, or to misuse \sysname to manipulate the overall system, e.g., by having benign devices identified as malicious by the system and sanctioned.
Subsequently, we will individually explain each goal and why it cannot be achieved by the adversary \adv.

\paragraph{Evade identification}
To evade the identification of nodes controlled by \adv, it can follow different strategies: (1)~try to prevent being detected initially, (2)~prevent being blamed, (3)~prevent that a consensus is found identifying the adversary-controlled node.

\mbox{  } \textit{Strategy 1:} 
To avoid initial identification \adv can (a)~stop interacting with the overall system and not participate in the integrity validation, or (b)~behave correctly according to the used integrity validation scheme used.
If \adv isolates itself while at the same time not answering to integrity validation request will ultimately lead to the conclusion that a node is not behaving correctly. 
However, given an appropriate integrity validation scheme \adv will not be able to pass it, unless it breaks the validation scheme, which is assumed to be secure.

\mbox{  } \textit{Strategy 2:} 
Once an adversary-controlled node has been recognized by another node, this node will send out a blame message to inform the network.
To prevent this, the adversary (a)~can compromise the blamer node, (b)~suppress the communication from the blamer node, or (c)~vilify the blamer node.

For (a), \adv would need to compromise the blamer \emph{before} it is able to send out the blame message. Although we consider this case out of scope (cf. \Cref{sec:advmodel}), even if \adv manages to compromise the blamer node, this node will be verified eventually and reported.
For (b), \adv first needs to continuously control \emph{all} communication channels of the blamer node; yet, even then \adv will be verified and reported eventually by any other benign node.
Lastly, in (c), \adv might try to discredit the blamer so other nodes will not believe the blame, i.e., the compromised node will broadcast a blame message accusing the blamer node.
In this situation, both nodes will be examined by the jury, which will uncover the real adversary. 

Hence, \adv only succeeds by entirely preventing the propagation of blame messages, i.e., \sysname is secure against the second attack strategy, as long as the assumptions hold that \adv does not have complete control over the network (cf. \Cref{sec:model}).

\mbox{  } \textit{Strategy 3:}
Finally, \adv can try to prevent that the network finds agreement regarding the compromise of a node.
The adversary can (a)~try to sabotage the election/forming of a jury, (b)~control a quorum of the jury, %
(c)~prevent interaction between jury members, or (d)~the broadcast of the jury decision.
Finally, (e)~\adv can distort the random jury election process to cause inconsistencies within the network that will affect the decision-making in the subsequent consensus phase.

To sabotage the jury election and forming, \adv needs to block overall communication in the network, which is assumed to not be possible (cf. \Cref{sec:advmodel}).
The adversary could also subvert the nodes to be part of the jury, e.g., to shut them off.
However, with high probability, a quorum of nodes will be elected that are not compromised by \adv, as long as the total adversary share does not exceed \advshare as we show in \Cref{sec:probs}.

In order to control a quorum of jury members, \adv can either compromise the jury members on-demand once they are elected.
This, however, requires \adv to be able to rapidly compromise a quorum of jurors, which contradicts our adversary model (cf. \Cref{sec:advmodel}).
Otherwise, \adv has to break the random jury election scheme to reliably get nodes that are under its control to be elected as jurors.
Since the jury is randomly selected, there is a chance that the adversary-controlled nodes get elected.
As we show in \Cref{sec:probs} this probability is negligible with the right choice of parameters for an adversary share up to \advshare.

To prevent the benign jurors from finding consensus, \adv can disrupt their communication.
First, \adv needs to prevent a quorum to actually disrupt the consensus.
Second, when preventing a quorum, the jury will trigger a re-election that results in a new jury.
To continuously disrupt each newly elected jury, \adv is required to disrupt any communication in the network; hence, contradicting our adversary model (cf. \Cref{sec:advmodel}).

Furthermore, \adv could try to prevent the jury from announcing the agreed-on result to the network, which implies that \adv needs to prevent all broadcasts by every juror.

Finally, \adv could try to manipulate the jury election process in order to prevent devices in the network to learn the correct list of jury members.
As a consequence, these devices would not accept the decision of the jury leading to inconsistencies between different nodes of the network.
However, this would require \adv to \emph{permanently} prevent the wait certificates by legitimate jury members from arriving at selected devices.
Given that the random jury election scheme does provide the guarantee that the elected jury is known to the entire network, all devices will eventually accept the decision made by a quorum of legitimate jurors as soon as they learn the list of legitimate jurors.
Even if some devices do not learn the decision of the jury, i.e., have differences in $JEL$ due to waiting certificates being withheld by \adv, this will have the effect of reducing the fault-tolerance of the subsequent Byzantine agreement, with `shortest $j$' nodes missing from a node's $JEL$ being replaced by other nodes from outside this set, essentially manifesting as an additional fault.
Thus, the security of \sysname depends on the security provided by the used schemes.

Hence, in order for \adv to succeed with strategy 3 it has to break one of the used schemes (integrity validation, random jury election, or consensus finding), has to prevent broadcast by all jurors, or be able to quickly compromise all jury members.
Each of these attacker capabilities violate our system and adversary model.

\paragraph{Manipulate system}
The adversary can also try to manipulate the system by misusing \sysname.
In particular, by blaming benign nodes \adv can try to get them sanctioned, e.g., excluded from the network to increase its own share of the network.
However, to achieve this \adv has to alter the integrity validation report of a benign node to convince the jury that the node is compromised.
This means \adv has to break 
the authentication method used by 
breaking a cryptographic primitive like signatures, which is not possible (cf. \Cref{sec:advmodel}).
Alternatively, \adv can aim to gain control over a decision-making majority of the jury to come to a malicious agreement that will be accepted by the entire network.
Here the same arguments hold as discussed above for strategy 3b and the probability of success is analyzed in the following (\Cref{sec:probs}). 

In summary, the adversary can only misuse \sysname when breaking one of the underlying schemes or with negligible probability, and thus fulfills the requirement~\ref{req:det_ident}.

\subsection{Probabilistic Analysis} \label{sec:probs}
The adversarial share of nodes \advshare \sysname can tolerate depends on the probabilities of electing compromised devices as jurors.
As the election is random, it may happen that enough adversarial nodes are elected for the Byzantine agreement to fail, as described in \Cref{sec:bftimpl}.
This section discusses the probabilities for different adversary shares \advshare as well as the chosen jury size $j$.
The joint decision is based on a Byzantine agreement, which means it fails if
more than $\lfloor (j-1)/3 \rfloor$ jurors are adversarial \cite{pbft}. 
While a larger $j$ reduces the chances of a failed election, \ac{bft} also induces a message complexity of $O(j^2)$. 
In \Cref{sec:eval_jury_size} we will evaluate this effect in our simulation.

If we have $n$ total nodes in our system, with $f$ of them being adversaries and
elect $j$ jurors, the probability of electing at least $\lfloor (j-1)/3 \rfloor$
adversarial nodes is:

\begin{equation} \label{eq:fail}
1-{{{j \choose {k+1}}{{n-j} \choose {f-k-1}}}\over {n \choose f}} \,_3F_2\!\!\left[\begin{array}{c}1,\ k+1-f,\ k+1-j \\ k+2,\ n+k+2-f-j\end{array};1\right]
\end{equation}

Where $k=\lfloor (j-1)/3\rfloor$ and $\,_pF_q$ is the generalized hypergeometric function. 
\Cref{eq:fail} is the cumulative distribution function of the hypergeometric distribution.

While this equation models the probability for the Byzantine agreement to fail, 
we can rectify a liveness violation by re-election.
In some applications, it may make sense to accept reduced fault-tolerance by increasing the quorum size required by PBFT from $\lfloor 2(j-1)/3 \rfloor+1$ to some greater value $q$. 
Then, $n-q \le \lfloor (j-1)/3 \rfloor$ faults are sufficient to cause a liveness violation, but a safety violation requires a greater number $2q-n$ of faults.
If the protocol reaches an impasse, another consensus round, including a new jury, is started that may succeed. 
This can be modelled as a Markov chain: we begin in an initial ``undecided''
state and transition to a ``success'' state if no more than $n-q$ adversarial nodes
are elected---guaranteeing agreement---and a ``failure'' state if at least $2q-n$ adversarial nodes are elected---allowing a safety violation.  
The failure state will eventually be reached with probability

\begin{equation} \label{eq:totalfail}
  \mathrm{P}[\text{Eventual Failure}] = \frac{\mathrm{P}[F \ge 2q-n]}{\mathrm{P}[F \ge 2q-j]+\mathrm{P}[F \le j-q]}
\end{equation}

and it will take on average $1/\mathrm{P}[j-q < F < 2q-j]$ elections to leave the ``undecided'' state.

\begin{figure}[ht!]
	\centering
	\includegraphics[width=\columnwidth,trim=0.2cm 0.3cm 0.2cm 0.35cm, clip]{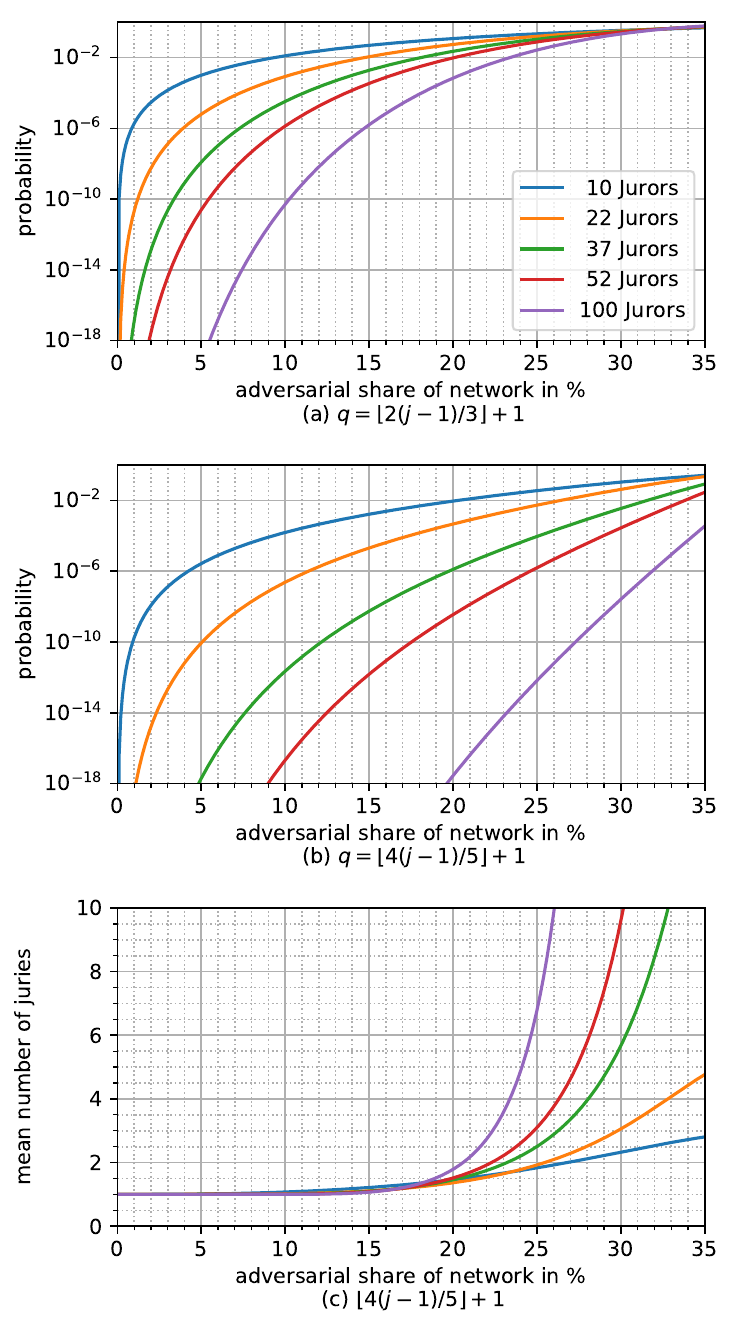}
	\caption{The probability of eventual safety violation of Byzantine agreement
		with a population size of $10\,000$
		given a threshold of (a) $q = \lfloor 2(j-1)/3\rfloor+1$ and (b) $q = \lfloor 4(j-1)/5\rfloor+1$, as well
		as the mean number of juries needed before agreement terminates, whether
		in success or \emph{total} failure, for $q = \lfloor 4(j-1)/5\rfloor+1$ in Figure (c). The
		distinctly colored graphs depict the probability development for different
		jury sizes $j$.  Note that the case depicted in (a) will always either terminate or suffer a
		safety violation with a single jury election, unlike that in (b) and (c)
		where several juries may be necessary.}
	\label{fig:eval_bftfail}
\end{figure}

Besides the threshold $q$, a primary factor affecting the probability of an
eventual safety violation is the jury size $j$. The more jurors are elected per
round, the lower the probability for the Byzantine agreement to fail. We
illustrate the influence of the jury size $j$ and BFT threshold $q$ in
\Cref{fig:eval_bftfail}. 
The choice of jury size $j$ and threshold $q$ is therefore
application-dependent, depending upon the appropriate trade-off between failure
probability, time to reach agreement, and performance.  We evaluate the latter of
these considerations in Section~\ref{sec:eval_jury_size}.

\section{Performance Evaluation}
\label{sec:eval_perf}
In this section, we measure the performance of our \sysname instance using a small-scale network running on real hardware, and use these results as the basis for simulating large-scale networks.
Regarding the results of these simulation campaigns, we first analyze the effects of differently chosen wait time parameters.
These parameters need to be chosen carefully to ensure the random election is consistent.
Afterwards, we examine the scalability of our \sysname instance for large networks regarding run-time and messaging overhead, showing sub-linear run-time growth in regards to network size. 
Finally, we show how this performance is affected by choosing different jury sizes.

\subsection{Prototype Measurements} \label{sec:impl_hw_eval}

For reference measurements to use in the large-scale simulation (see \Cref{sec:eval_sim}) we deployed our \sysname implementation (see \Cref{sec:impl2}) on a setup of ten nodes.
We used Raspberry Pi 3 Model B+~\cite{raspberry} as the platform and connected a distance sensor for attestation.
This platform is comparable to ARM's line-up of chips specifically designed for vehicles~\cite{arm_cars} in terms of computational power and capabilities.
The Raspberries are running a Raspbian Linux after OP-TEE is loaded. 
We connected all nodes to a router via WiFi for communication between them.

\noindent
\textbf{Enhanced attestation.}
To demonstrate that {\sysname} can scale with a more complex integrity validation scheme as well, we also consider Data Integrity Attestation (DIAT)~\cite{diat}, which targets trustworthy data exchange for collaborative autonomous networks, such as cars or drones.
DIAT is a remote attestation approach that can detect even sophisticated software attacks, such as run-time attacks~\cite{FC08}.
Due to the increased complexity, it also needs more processing for both the attestation generation and validation.

\begin{table}[tb]
	\footnotesize
	\caption{The measured run-times of individual processing steps.} \label{tbl:runtime} 
	\begin{tabularx}{\columnwidth}{|Y|Y|Y|Y|}
		\hline
		wait certificate generation & static attestation generation & static attestation validation & BFT process + Schnorr-sign \\ \hline
			42 ms & 166 ms & 1 ms & 14 ms \\ \hline
	\end{tabularx}
	\begin{tabularx}{\columnwidth}{|Y|Y|}
		\hline
		DIAT GPS attestation generation & DIAT GPS attestation validation  \\ \hline
		835 ms & 849 ms \\ \hline
	\end{tabularx}
\end{table}

The top half of \Cref{tbl:runtime} shows our measurements regarding run-time of our implementation.
The \ac{bft} and Schnorr signing is fluctuating depending on jury size, so for the simulation, described in \Cref{sec:eval_sim}, we chose to use the worst-case (14 ms). 
The bottom half of \Cref{tbl:runtime} shows the \acs{diat} run-time numbers for attesting a GPS module, as reported in the paper~\cite{diat}.

\subsection{Simulation} \label{sec:eval_sim}
To evaluate the performance of \sysname for large numbers of devices, we used the OMNeT++ network simulator \cite{omnetpp}. 
We implemented \sysname at the application layer and used the measurements described in \Cref{sec:impl_hw_eval} to set the processing times for the individual steps taken by each node. 

Our network is configured in a square mesh topology, with roughly the same height and width. 
Every node has four links to its neighbors, except the nodes at the edge of the network.
To make the simulation representative, we used dynamic communication delays between nodes.
We measured the latencies in our distributed setup described in \Cref{sec:impl_hw_eval} in different scenarios, such as highly varying distances commonly encountered in vehicle-to-vehicle communication.
We measured 3 ms at best and 78 ms at worst for one-way delays.
For the simulation, each communication link gets a random delay assigned between these two measurements.
Routing for dynamic networks is an orthogonal problem~\cite{vanet,dtn} and does not contribute to a meaningful evaluation of \sysname.
Thus, we use a simple on-demand routing algorithm.

We evaluate \sysname for different network sizes, from $1\,000$ to $100\,000$ nodes.
We also split measurements into the different phases. 
We simulate the first round of our \sysname instance with the following phases\footnote{Subsequent rounds will be faster, as no election is needed.}:

\begin{enumerate}
	\item Initial Attestation: Generation of the initial attestation report $\sigma$ by the blamed node and the integrity validation of $\sigma$ by the blamer.
	\item Blame: Broadcasting the initial blame message.
	\item Election: The election process to elect $j$ jurors.
	\item \ac{bft}: The consensus protocol, including the validation of $\sigma$ by each juror.
	\item Decision: Broadcasting the outcome of the \ac{bft}.
\end{enumerate}

Notice that this represents the worst case, i.e., the upper bound regarding run-time and message overhead, as it includes both the election and the complete \ac{bft}.
Further, to minimize variation of individual simulation runs, due to the random nature of our scheme, we average every individual parameter configuration over 100 runs with different random numbers.

\subsection{Election Wait Time}
\label{sec:eval_waittime}

We evaluate the time parameters $t_{max}$ and $t_{ele}$, as they contribute significantly to the performance characteristics of \sysname. 
$t_{max}$ is the maximum wait time regarding the randomly chosen wait time for each node.
After a node is done waiting, it will wait an additional time $t_{ele}$ while collecting other waiting numbers.
Afterwards, it will assume the election to be mostly complete, i.e., to have a mostly matching $JEL$.
If $t_{ele}$ is chosen very small, the election itself will be faster; however, the individual $JEL$s may also be still inconsistent among the nodes. 
To measure this effect we executed a parameter study for differently chosen time parameters with $n=2\,000$, $j=22$ and $t_{min}=100ms$.

\begin{figure}[t!]
	\centering
	\includegraphics[width=\columnwidth,trim=0.2cm 0.3cm 0.35cm 0cm, clip]{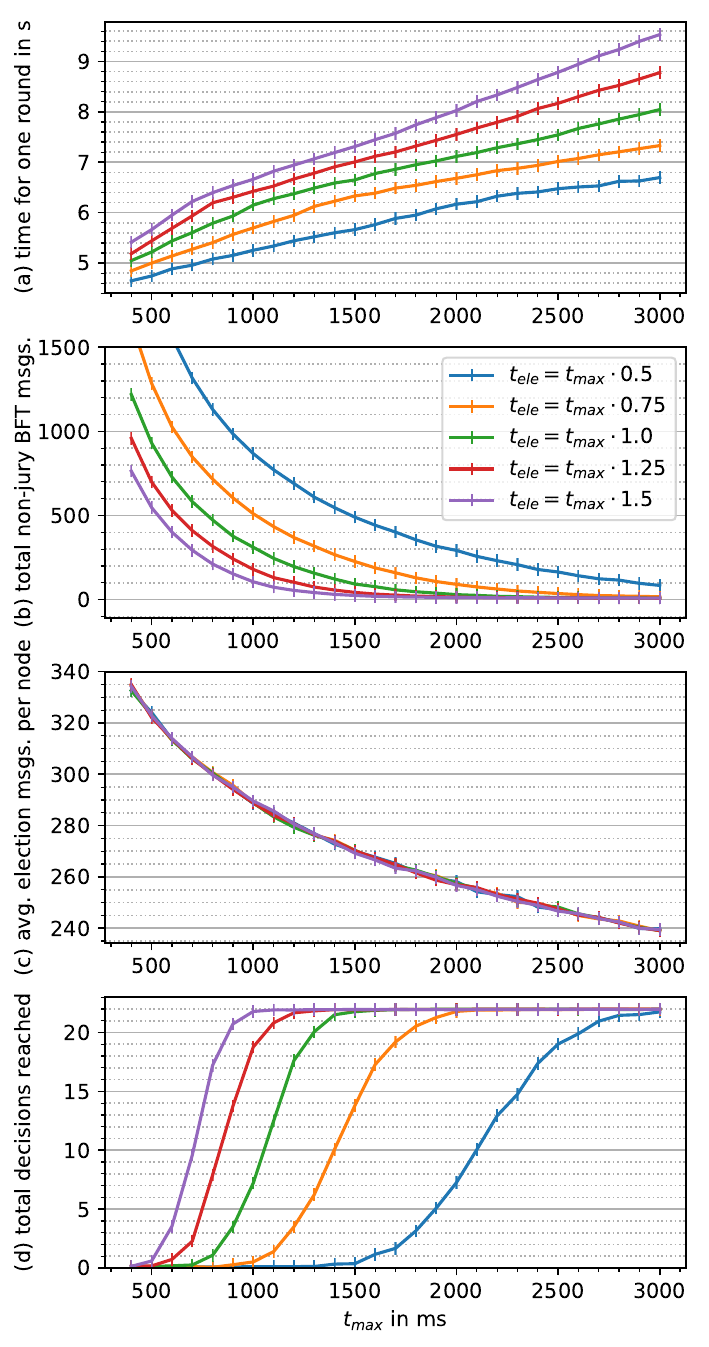}
	\caption{(a) The time for one round, (b) the total number of non-juror \ac{bft} messages, (c) the average number of election messages per node and (d) the total number of reached decisions, all for differently chosen $t_{max}$ and $t_{ele}$. Simulated with $n=2\,000, j=22$, $t_{min}=100ms$.}
	\label{fig:eval_waittime}
\end{figure}

\Cref{fig:eval_waittime} shows the results. 
In (a) we can see the effect on the execution time of one round. 
It is primarily tied to $t_{ele}$, as can be seen if comparing different $t_{max}$ that result in the same $t_{ele}$.
The graph (b)  in turn shows how many unjustified \ac{bft} messages were received. 
While a lower $t_{ele}$ reduces the execution time, it also increases the number of nodes falsely assuming to be jurors.

\begin{figure}[ht]
	\centering
	\includegraphics[width=\columnwidth,trim=0.2cm 0.3cm 0.3cm 0.25cm, clip]{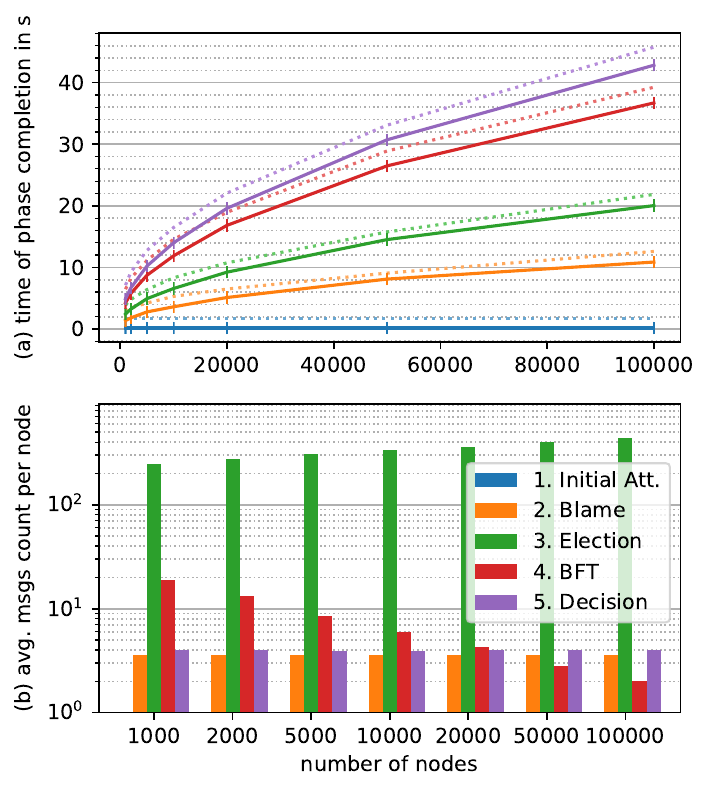}
	\vspace{-0.3cm}
	\caption{(a) The completion time of each phase, results for static attestation and DIAT (dotted lines), and (b) the average number of
    messages sent per node, all split into the individual phases for increasing $n$. Simulated with $j=22$.}
	\label{fig:eval_phases}
\end{figure}

\Cref{fig:eval_waittime} (c) shows the average number of messages per node for the election phase. 
This shows how many nodes actively participate in the election, i.e., nodes assuming their wait time still has merit after waiting for $t_{ele}$.
However, (d) shows how many jurors get to the point of sending out a decision. 
This measurement should ideally match the chosen $j$, so 22 in this case.
Even though lower numbers would suffice, matching $j$ allows the network to better account for other faults in the consensus phase.
Yet, it can be seen that if the time parameters are chosen too small, not the entire jury can reach a decision or none at all, as the nodes' $JEL$ will diverge to the point where no quorum can be established among the jury.

With these measurements in mind, we chose a time parameter configuration for our further evaluation, keeping the discovered trade-offs in mind: $t_{max}=1\,000ms$ and $t_{ele}=1\,500ms$ for $n=2\,000$.
After many experiments for different network sizes, we found a dynamic configuration, which works for all network sizes.
Based on the average delay of 37.5 ms between nodes, we set $t_{ele}=\sqrt{n} \cdot 37.5ms \cdot 0.9$ and $t_{max}=t_{ele} \cdot \frac{2}{3}$.
The square root of $n$ times the delay reflects half of the worst-case communication delay in our topology and 0.9 means it can be 10\% lower than that, while still resulting in a reliable election.

\subsection{Per-Phase Performance for Large Networks} \label{sec:eval_phases}
In this section, we examine the \emph{Efficiency} and \emph{Scalability} of \sysname, two main requirements (cf. \Cref{sec:requ}). 
\Cref{fig:eval_phases} (a) shows the run-time measurements. 
Note that the measurements per phase are denoted as the \emph{absolute} simulation time at the last processed message of the respective phase---phases overlap as progress is made in parallel. 
The top purple line represents the time of the last received decision message in the network, and thus the total time for one entire \sysname round.
We consider two attestation schemes with differing complexity (see \Cref{sec:impl_hw_eval}). One regarding our static attestation and the other regarding the more complex DIAT~\cite{diat} (shown as dotted lines).
A network of $n=100\,000$ takes $42.83s$ with our static attestation approach and $45.76s$ with DIAT, fulfilling requirement~\ref{req:scalability}.
Note that the worst-case communication delay between any two nodes in this scenario is $23.71s$ on average.

A naive and simplified solution to the problem would be to let all devices attest every other device individually. 
The time this case takes for $n$ nodes is the combined time of the generation and verification of an attestation multiplied by $(n-1)$.
This does not consider communication delay and assumes perfect parallelization between the nodes. 
This naive case would take almost 3 minutes for $n=1\,000$ ($\sim$14 minutes with DIAT) and over $4$ hours for $n=100\,000$ ($\sim$23 hours with DIAT); hence, the naive approach is $\sim$390x slower ($\sim$1800x slower with DIAT) than \sysname, and thus we deem \sysname efficient (\ref{req:efficieny}).

\Cref{fig:eval_phases} (a) also shows the individual measurements per phase.
The third line in green shows how long the election takes. 
The time for the election overlaps with the blame broadcast, implying that the election requires the most time of the scheme. 
The red line, second from the top, shows when the \ac{bft} is finished.
\Cref{fig:eval_phases} (b) show the overhead per phase in terms of message count. 
Note that we consider all messages for these measurements, i.e., including forwarded messages by nodes in between the route.
The graph reveals that the election phase (green bars second from the left respectively) generates the most overhead.
The total message overhead for $n=100\,000$ is 333.01 messages per node. 
However, assuming a subsequent round with a jury already in place, the total message overhead without the election phase is reduced to 9.63 messages per node. 

\subsection{Jury Size} \label{sec:eval_jury_size}
\begin{figure}[t]
	\centering
	\includegraphics[width=\columnwidth,trim=0.25cm 0.3cm 0.35cm 0.35cm, clip]{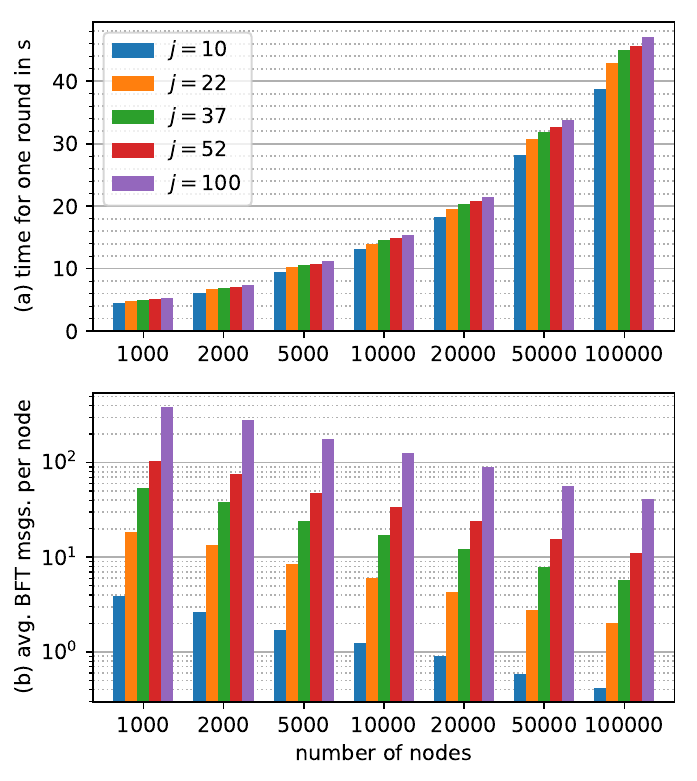}
	\vspace{-0.3cm}
	\caption{(a) The time for one round and (b) the average number of \ac{bft} messages sent per node for different $j$.}
	\label{fig:eval_jurysize}
\end{figure}

The following examines the effects of differently chosen jury sizes $j$ on our instance of \sysname. 
\Cref{fig:eval_jurysize} (a) shows the run-time for one round. 
For large juries, like $j=100$, in a large network, like $n=100\,000$, the difference on the run-time compared to $j=10$ is $8.3s$ (or 21.4\%). 
This is due to the individual \ac{bft} steps being able to execute in parallel. 

The second graph (b) show the average message count per node of the \ac{bft} phase.
The $O(n^2)$ message complexity for two \ac{bft} phases are apparent.
Nevertheless, the closest case we could find to compare the election overhead against the \ac{bft} overhead is $n=1\,000$ and $j=100$. 
Here the average message overhead per node for the election is 590.65 against the 387.11 for the \ac{bft} message overhead. 
Thus, a \ac{bft} round is more efficient than an election in overall terms. 
However, \ac{bft} also concentrates the overhead on the jurors and the routes between them, compared to the more uniformly distributed overhead by the election.


\section{Discussion} \label{sec:discussion}
\acresetall
This section discusses possible extensions to \sysname.

\paragraph{Broadcast}
To reduce communication costs in \sysname, a gossip protocol can be used.
Gossip protocols randomly send broadcast messages to a set number of neighbors, which in turn do the same~\cite{gossip}. 
These protocols are probabilistic in nature, yet, perform well on average and significantly reduce overhead compared to flooding~\cite{gossip}.

\paragraph{Monitoring}
An aspect that could be changed is the on-demand nature of the initial integrity validation.
For example, one could have all nodes regularly check all their neighbors instead.
These validations can be entirely local per node and the \sysname process would only be triggered when a node actually notices inconsistencies.
This way, every node would be validated eventually, so even a passive adversary cannot hide.

\paragraph{Dynamic Jury}
The jury size does not have to be fixed over the life span of a \sysname instance. 
It might be advantageous to dynamically adjust the jury size when required, e.g., increase the jury size when many blames occur in a short time frame. 
This would dynamically adjust the security probabilities along the network's needs at the time.

\paragraph{Jury Decision}
Part of a practical instantiation of \sysname is the resulting reaction of the jury to a confirmed adversarial node. 
This by itself is a vastly complex topic and highly dependent on the use case. 
Straightforward expulsion of the malicious node as a result of the jury decision is not possible in many use cases, e.g., cyber-physical systems like autonomous cars, where expulsion from the network does not prevent them from affecting the system.
In these situations, some other response might be more appropriate. 
Mechanisms exist for \emph{self-healing}~\cite{selfheal}, in which a faulty node is returned to a valid state.
In this case, we might choose to use \sysname not to exclude an adversarial node, but to decide whether a node will be \emph{added} to the network.


\section{Related Work} \label{sec:related}
\paragraph{Distributed Outlier Detection}
In the field of Wireless Sensor Networks (WSNs), outlier detection is used to identify unusual sensor readings to detect faults, exceptional events, or malicious attacks.
There are different techniques to identify outliers, from simple statistical methods (e.g., Gaussian-based models), up to classification-based methods, such as machine learning~\cite{wsn_survey}.
The outlier evaluation can be done locally per node or by a central node trusted to handle the decision making~\cite{wsn_survey}.
To distribute this task among the nodes, one proposal is for each node to only consider the neighbors for outliers and keep exchanging decisions among the network until a global view is achieved~\cite{wsn_neigbors}.
An efficiency improvement to this approach is to do a majority vote among a neighborhood, which can be exchanged with other neighborhoods~\cite{wsn_vote}.

However, these distributed approaches require nodes to exchange a significant amount of data and messages to converge to a global view.
This limits their applicability to large-scale networks, unlike \sysname which easily supports $100\,000$ nodes.

\paragraph{Distributed Intrusion Detection}
Similar to outlier detection, Distributed Intrusion Detection Systems find anomalies specifically for network traffic in WSNs~\cite{ids_majority_wsn,ids_central} and mobile ad-hoc networks~\cite{ids_majority_manet,ids_cluster1,ids_cluster2,ids_cluster_hierarchy}.
These approaches also use similar techniques to identify anomalies; yet, there are also approaches based on prior knowledge of normal operations or a defined specification~\cite{ids_adhoc_surv}.
There are different ways to distribute data aggregation and anomaly processing to reduce communication overhead.
However, in the context of this work, the key aspect is how decisions are made.
One approach is for nodes to collaborate with their neighbors for the measurements and decide via majority vote on a suspected node~\cite{ids_majority_manet,ids_majority_wsn}.
This assumes compromised devices are in the minority in every neighborhood, as colluding adversaries may easily form local majorities otherwise.
Another method is to separate the network into clusters and have their members elect the \emph{cluster heads}, which representatively make the decisions~\cite{ids_cluster1,ids_cluster2}.
However, either the election is disregarding that it might be malicious~\cite{ids_cluster1}, or the assumption is a low threat environment implying a low probability of electing a malicious node~\cite{ids_cluster2}.

In contrast, \sysname can tolerate many malicious devices in any distribution.
Other works in this field introduce some form of centralization, such as a trusted base station making decisions~\cite{ids_central, ids_central2}, or a privileged group of nodes on top of a hierarchy of clusters~\cite{ids_cluster_hierarchy}.

\paragraph{Collective Attestation}
The first step towards scalable attestation of large groups of interconnected devices, i.e., collective attestation, was made by SEDA~\cite{seda}. 
SEDA, like all other schemes that followed in the collective attestation literature~\cite{sana,seed,wise,salad} assume a central verifier; hence, they are not applicable in the autonomous scenarios targeted by \sysname.
Further, these approaches aim to verify the whole system at once on request, whereas \sysname ensures security in a sustainable way by giving all nodes a tool to continuously and autonomously validate each other.


\section{Conclusion} \label{sec:concl} 
In this work, we presented \sysname, the first scheme to efficiently identify adversaries in large networks of autonomous collaborating devices.
\sysname combines random elections, consensus and integrity validation methods in a flexible scheme, where each of these components are interchangeable.
We have demonstrated the scalability of an exemplary instance as well as provided the basis to construct use-case specific instances of \sysname.

\begin{acks}
	We thank N.~Asokan (University of Waterloo) for his useful feedback. This work has been supported by the German Research Foundation (DFG) as part of the project S2 within the CRC 1119 CROSSING, by the ASSURED project funded by the EU's Horizon 2020 programme under Grant Agreement number 952697, by German Federal Ministry of Education and Research (BMBF) within the project iBlockchain, by the Academy of Finland (grant 309195), by the European Space Operations Centre with the Networking/Partnering Initiative, and by the Intel Collaborative Research Institute for Collaborative Autonomous \& Resilient Systems (ICRI-CARS).
\end{acks}

{
	\bibliographystyle{ACM-Reference-Format}
	\bibliography{SADAN}
}


\appendix
\section{Appendix}
This section will give a thorough discussion on the identified options mentioned in \Cref{sec:framework}.

\subsection{Integrity Validation Options} \label{sec:options:validation}

\noindent\paragraph{Outlier Detection}
Unsupervised outlier detection for sensor data \cite{wsn_survey,wsn_vote,wsn_neigbors}, which is the prevalent in Wireless Sensor Networks, enables the validation of measurements reported by individual nodes.
Hence, it allows to identify malicious nodes sending manipulated data. 
We outline the different types of anomalies and detection techniques in \Cref{sec:related}.
The detected outliers as well as the required accompanying data (e.g., outlier reports of the blaming node's neighbors~\cite{wsn_neigbors}) can be used as the disputed evidence in \sysname.

\noindent\paragraph{Intrusion Detection}
Intrusion Detection Systems (IDS) monitor for anomalies in network traffic in order to discover intrusions. 
In \sysname, detected anomalies can serve as evidence. 
Instead of sending an \emph{attack report}~\cite{ids_central2} or aggregated observations~\cite{ids_central} to a central authority, this data can be used as evidence to trigger an investigation by a jury.
\Cref{sec:related} outlines several distributed approaches for IDS proposed in the literature.

\subsection{Random Election Options} \label{sec:options:election}
\noindent\paragraph{Algorand} \label{sec:algorand}
In Algorand~\cite{algorand} a delegation group is randomly elected to propose new blocks. 
Here, each node draws a number based on the Verifiable Randomness Function (VRF). 
The lowest numbers win the election, and thus the delegation group is elected. 
The VRF works as a deterministic source of randomness and as such is publicly verifiable. 
Put simply, each node's individual random number is the hash of the concatenation of its identity, i.e., public key, and the last block's hash. 
This results in a random number that is verifiable by all participants, as only public information is necessary to calculate it.
Algorand also needs to protect against Sybil attacks, as nodes may freely join the network.
Each node has stake, i.e., the amount of money they own in the system, which is used to assign weight to their random number.
Thus, the bigger a node's monetary stake the higher the chances to be elected and vice versa.
However, if all participants are known, the election itself can be executed without requiring stake.
It also implies a large message overhead, as all participants broadcast their election numbers virtually simultaneously.
Further, Algrorand builds on a adapted binary consensus algorithm adapted to transaction block selection, and thus is specifically designed for cryptocurrencies.

\paragraph{Byzcoin}
For selecting the consensus group Byzcoin~\cite{byzcoin} requires to mine consensus blocks via Proof-of-Work (PoW). 
Then, a chosen number of the last successful miners emerges as the group executing \ac{bft}.
This elected group will then propose a new transaction block together.
The key issue when using PoW is that in a heterogeneous network some less powerful nodes have a significant disadvantage in the election.
Further, an adversary may even use a powerful external machine to exceed the processing power of the entire network.

\paragraph{Deterministic Random Jury}
Another simplified approach is to use a single verifiable source of randomness instead of many. 
Thus, the drawn number elects the whole jury. 
This way, instead of having all nodes announce their number individually, the whole network would deterministically know who is part of the next jury. 
For example, a counter of the \sysname round could be concatenated with all of the identities of the previous jury and subsequently hashed as the source of randomness. 
This approach has a very low overhead for the election phase; yet, one has to consider the possibility that an elected juror crashed, which leads to additional faults in the consensus phase.

\subsection{Consensus Options} \label{sec:options:consensus}

\noindent\paragraph{Simple Majority}
If we assume to elect a new jury every round and every juror has a random number, we can extract an inherent order of requests. 
On two conflicting requests, there will be two separate elections with two separate juries. 
In such a case, the juries decide which request is executed first by comparing their election results.
With the order being ensured, a simple majority vote among the jury suffices.
While this requires little overhead for the consensus phase, it requires a new election each round.

\paragraph{\ac{bft} with Enhancements}
In \Cref{sec:probs} we show that with increasing jury size, the probability to fail decreases significantly. 
However, a larger jury also implies a larger overhead due to the $O(n^2)$ message overhead of \ac{bft}~\cite{pbft}.
To counter this, different enhancements of \ac{pbft} can be employed to reduce  complexity. 
For example, the speculative case \cite{zyzzyva}, skipping \ac{pbft} phases, or the optimistic case \cite{rebft}, halving the consensus group.
However, both reduce overhead only for the benign case.
Thus, if we expect Byzantine events to be rare, it may be feasible to consider larger jury sizes with inherently better security guarantees. 
Another approach is to involve a message aggregation scheme \cite{byzcoin} to significantly reduce message complexity. 
If we have a trusted component available, we can also employ a trusted monotonic counter, which removes the need for \acs{bft}'s prepare phase entirely as well as reducing the required quorum to half plus one nodes \cite{minbft, fastbft}.

\end{document}